\documentclass[eqsecnum,secnumarabic,preprint,aps,prd,preprint,tightenlines,
endfloats,draft,showpacs,nofootinbib,nobibnotes]{revtex4}

\usepackage{latexsym,amsfonts,amssymb,psfig}

\begin{document}

\title{Dynamical Casimir Effect and Quantum Cosmology}
\author{I. Brevik}
\email{iver.h.brevik@mtf.ntnu.no}
\affiliation{Division of Applied Mechanics, Norwegian University of
Science and Technology, N-7491 Trondheim, Norway}

\author{K. A. Milton}
\email{milton@mail.nhn.ou.edu}
\homepage{http://www.nhn.ou.edu/
\affiliation{Department of Physics and Astronomy, The University of
Oklahoma, Norman, OK 73019, USA}

\author{S.D. Odintsov}
\email{odintsov@mail.tomsknet.ru, odintsov@ifug5.ugto.mx}
\altaffiliation[Present address:  ]{Instituto de F\'\i 
sica de la Universidad de Guanajuato,
37150 Le\'on, M\'exico}
\affiliation{Department of Mathematics and Physics, Tomsk State
Pedagogical University, Tomsk 634041, Russia}

\author{K.E. Osetrin}
\email{osetrin@tspu.edu.ru}
\affiliation{Department of Mathematics and Physics, Tomsk State Pedagogical 
University, Tomsk 634041, Russia}

\date{\today}
\preprint{OKHEP--00--01}

\begin{abstract}
We apply the background field method and the effective action formalism to
describe the four-dimensional dynamical Casimir effect. Our picture
corresponds  to the consideration of quantum cosmology for
an expanding FRW universe (the boundary conditions act as a moving mirror) 
filled by a quantum massless GUT which is
conformally invariant. We consider cases in which the static
Casimir energy is  attractive and repulsive.  Inserting the simplest possible
inertial term, we find, in the adiabatic (and semiclassical) approximation,
the dynamical evolution of the scale factor and the dynamical Casimir 
stress analytically and numerically (for $SU(2)$ super Yang-Mills theory).
Alternative kinetic energy terms are explored in the Appendix.
\end{abstract}


\maketitle
\section{Introduction}

The Casimir effect \cite{casimir} can be regarded as the change in the
zero-point fluctuations due to  nontrivial boundary conditions. Surveys of
the effect are given, for instance, by Plunien et al.~\cite{plunien86},
Mostepanenko and Trunov \cite{mostepanenko97}, and Milton \cite{milton99}.
The recent ``resource letter" of Lamoreaux \cite{lamoreaux99} contains a
wealth of references, although it is admittedly highly incomplete.

In the past, the Casimir effect has been considered as a {\it static} effect.
Growing interest in recent years has been drawn to the {\it dynamical}
variant of the effect, meaning, in essence, that not only the geometrical
configurations of the external boundaries (such as plates) but also their
velocities play a physical role. Moore \cite{moore70} is probably the first
to have considered the dynamical Casimir effect. Examples of more recent
references are \cite{sassaroli94} and \cite{dodonov89}.

The recent paper of Nagatani and Shigetomi \cite{nagatani99} is an
interesting development in this direction. These authors focused attention
on the fact that if moving boundaries (mirrors) create radiation, the
mirrors have to experience a reaction force. They proposed an effective
theory for the back reaction of the dynamical Casimir effect in (1+1)
dimensions for a scalar field, this theory being constructed by the background
field method in the path integral formalism.
In fact, they considered a kind of $2d$ quantum cosmology for describing the
dynamical Casimir effect. 

In the present paper we show how to apply the effective action formalism,
using the background field method, to formulate the dynamical Casimir effect 
in four dimensions in a convenient and elegant form. We are able to
consider an arbitrary matter content (typically a 
grand unified theory or GUT) and present the dynamical
Casimir effect as a kind of quantum cosmological model.  Using the background
field method, we treat the geometrical configuration of the boundaries
classically, but consider the GUT in the interior region as a quantum object.

In the next section we consider a GUT in a three-dimensional space, where
the size of the space $a(t)$ is a dynamical variable. Similarly to
\cite{nagatani99} we make use of the adiabatic approximation. Exploiting
the conformal invariance of the theory we calculate the anomaly-induced
effective action $W$. In the simplest case (a torus), $W$ is given by 
Eq.~(\ref{2.4}). We consider the static Casimir energy in section 3, 
and show that, for the usual boundary conditions on the torus, the Casimir
energy is attractive.
In section 4 we start with the effective action, Eq.~(\ref{4.1}), for 
the dynamical case.
Introducing a mass $m$ associated with the scale factor $a$, with a
corresponding kinetic energy in the low-velocity approximation equal to
$\frac{1}{2}m\dot{a}^2$ (a phenomenological term), we then consider
two cases.  If the Casimir energy is attractive, we derive in Eq.~(\ref{4.15})
the time variation of $a(t)$ for large values of $t$, the last term in the
expression being a (special case of the)
dynamical correction to the pure quasistatic Casimir result. For the perhaps 
less realistic repulsive case, the small and large time behavior
of the Casimir behavior is extracted in Eqs.~(\ref{larget}), (\ref{smallt}).  
Numerical results in both cases
are given in Section 5.  The behavior of the scale
factor in the two cases is shown in Figs.~\ref{fig1} and \ref{fig3},
while the dynamical  stress on the torus is presented in Figs.~\ref{fig2}
and \ref{fig4}.  In the Appendix we discuss the effects of alternative
kinetic energy terms.

\section{The effective action}

Let us consider conformally invariant, massless matter in $4d$-dimensional
space-time. The matter may correspond to some GUT (say, $SU(5)$, $SO(10)$, 
or any
other alternative). We are interested first in the study of the static Casimir
effect for such a theory when the field is assumed to be bounded in a
three-dimensional region. In other words, we are interested in a space having
the form ${\mathbb R}^1 \otimes K^3$, where as $K^3$ one can take any manifold
permitting an
exact Casimir effect calculation. It can be ${\mathbb S}^3,\, {\mathbb T}^3,\,
{\mathbb S}^1 \otimes {\mathbb S}^2$
or any other compact manifold with a known spectrum of the d'Alembertian
operator. We limit ourselves to ${\mathbb T}^3$ or ${\mathbb S}^3$, for the sake of
simplicity.

Suppose now that our GUT lives in such a three-dimensional space, where the
size of the space is a dynamical variable (moving mirror or moving
universe). Hence, we will be interested in the
dynamical Casimir effect in a three-dimensional region and the back-reaction
from the induced radiation on the moving background geometry.
 We shall use the adiabatic approximation in this
study. A great simplification comes from adopting  a physical picture in which
the Casimir effect is described as  an effective action in curved spacetime
(see
\cite{BOS} for an introduction). Here, spacetime is taken to be an expanding
universe with topology ${\mathbb R}^1 \otimes {\mathbb S}^3$ or
${\mathbb R}^1\otimes {\mathbb T}^3$. The corresponding metric is given by
\begin{eqnarray}
ds^2=dt^2-a^2(t)ds_3^2,
\label{2.1}
\end{eqnarray}
where $ds_3^2=dx^2+dy^2+dz^2$ for ${\mathbb T}^3$ (coordinates are restricted
by all radii being equal), or the line element of a three-dimensional sphere
${\mathbb S}^3$.

Let us calculate now the effective action for such a GUT. Using the fact that
the theory is conformally invariant we may use the anomaly-induced effective
action \cite{reigert84}:
\begin{eqnarray}
W &=& b \int d^4 x \sqrt{-\bar{g}} \bar{F}\sigma\nonumber\\
&&\mbox{}+b'\int d^4 x
\sqrt{-\bar{g}}\; \{ \sigma [ 2 \bar{\Box^{\vphantom{2}}}^2
+ 4{\bar{R}}^{\mu\nu}{\bar{\nabla}}_\mu
{\bar{\nabla}}_\nu-\frac{4}{3}\bar{R}\,\bar{\Box^{\vphantom{2}}}
+\frac{2}{3}({\bar{\nabla}}^\mu \bar{R})({\bar{\nabla}}_\mu) 
]\sigma
+ (\bar{G}-\frac{2}{3} \bar{\Box^{\vphantom{2}}}\,\bar{R})\sigma \}
\nonumber\\
&&\mbox{}-\frac{1}{12}\times\frac{2}{3}(b+b')\int d^4x
\sqrt{-\bar{g}}
[\bar{R}-6\bar{\Box^{\vphantom{2}}}\sigma
-6({\bar{\nabla}}_\mu \sigma)({\bar{\nabla}}^\mu \sigma) ]^2,
\label{2.2}
\end{eqnarray}
where our metric is presented in conformal form. Thus
${\rm g}_{\mu\nu}=e^{2\sigma}\bar{g}_{\mu\nu}$,  $\sigma=\ln a(\eta), ~\eta$
is the
conformal time, $F$ is the square of the Weyl tensor, and $G$ is the square of
the Gauss-Bonnet invariant. Overbar quantities indicate that the
calculation is
made with $\bar{{\rm g}}_{\mu\nu}$. Further,
\begin{eqnarray}
b&=&\frac{1}{120(4\pi)^2}(N_0+6N_{1/2}+12N_1),\nonumber\\
b'&=&-\frac{1}{360(4\pi)^2}(N_0+11N_{1/2}+62N_1),
\label{2.3}
\end{eqnarray}
where $N_0$, $N_{1/2}$, and  $N_1$ are the numbers of 
scalars, spinors, and vectors. For
example, for ${\cal N}=4$  $SU(N)$ super YM one gets \cite{brevik99}
$b=-b'=(N^2-1)\left[4(4\pi)^2\right]^{-1}$.  We also adopt the scheme
wherein the $b''$-coefficient of the $\Box R$ term in the conformal anomaly
is zero. Being ambiguous, it does not influence the dynamics \cite{brevik99}.

As the simplest case we consider henceforth a torus. Then
\begin{eqnarray}
W=\int d\eta\,\left[ 2b'\sigma\sigma''''-2(b+b')(\sigma''+\sigma^{\prime2})^2
\right].
\label{2.4}
\end{eqnarray}
 This is a typical effective action for a GUT in a 
Friedman-Robertson-Walker (FRW) Universe of a special form.

\section{The Static Casimir energy}

Let us briefly overview the static Casimir effect for a torus 
of side $L$ (for more detail, see \cite{EOR}).
The Casimir energies associated with massless spin-$j$ fields is
\begin{eqnarray}
{\cal E}_{N_j}=\frac{N_j}{2}(-1)^{2j}
\sum_{{\bf n}\in{\mathbb Z}^3}\omega_{{\bf n},j},
\label{3.1}
\end{eqnarray}
where we see the appearance of the characteristic minus
sign associated with a closed Fermion loop.  The frequency of each mode
is given by
\begin{eqnarray}
\omega_{{\bf n},j}^2=\left(\frac{2\pi}{L}\right)^2\sum_{i=1}^3
(n_i+{\rm g}_i^{(j)})^2,
\,\,\,\, {\bf n}=(n_1,\,\,n_2,\,\,n_3),
\label{3.2}
\end{eqnarray}
Here ${\rm g}_i^{(j)}=0, 1/2$ depending on the
field type chosen in ${\mathbb R}^1\otimes{\mathbb T}^3$.

We use the $p$-dimensional Epstein zeta function
$Z_p\left| \begin{array}{ccc}
{\rm g}_1,&\dots,&{\rm g}_p\\
h_1,&\dots,&h_p 
\end{array}\right|(s)$ defined for $\Re \,s>1$ by the formula
\begin{eqnarray}
Z_p\left| \begin{array}{ccc}
{\rm g}_1,&\dots,&{\rm g}_p\\
h_1,&\dots,&h_p 
\end{array}\right|(s)&=&
\sum_{{\bf n}\in {\mathbb Z}^p}{'}\left[(n_1+{\rm g}_1)^2+...
+(n_p+{\rm g}_p)^2\right]^{-ps/2}\nonumber\\
&&\times \exp\left[2\pi i(n_1h_1+...+n_ph_p)\right],
\label{3.3}
\end{eqnarray}
where ${\rm g}_i$ and $h_i$ are real numbers, and the prime means omitting
the term with $(n_1,...,n_p)=(-{\rm g}_1,...,-{\rm g}_p)$ if all the ${\rm
g}_i$
are integers. For $\Re\,s<1$ the Epstein function is understood to be the
analytic continuation of the right-hand side of Eq.~(\ref{3.3}). 
Defined in such a way,
the Epstein zeta function obeys the functional equation
\begin{eqnarray}
&&\pi^{-ps/2}\Gamma\left(\frac{1}{2}ps\right)
Z_p\left| \begin{array}{ccc}
{\rm g}_1,&\dots,&{\rm g}_p\\
h_1,&\dots,&h_p \\
\end{array}\right|(s)
=\pi^{-p(1-s)/2}\Gamma\left(\frac{1}{2}p(1-s)\right)
\nonumber\\&&
\times \exp\left[-2\pi i({\rm g}_1h_1+...+{\rm g}_ph_p)\right]
Z_p\left| \begin{array}{ccc}
h_1,&\dots,&h_p \\
-{\rm g}_1,&\dots,&-{\rm g}_p
\end{array}\right|(1-s),
\label{3.4}
\end{eqnarray}
The function (\ref{3.3}) is an entire function in the complex $s$ plane except
for the case when all $h_i$ are integers. In the latter case the function
(\ref{3.3})
has a simple pole at $s=1$.

Using Eq.~(\ref{3.3}) we have
\begin{eqnarray}
{\cal E}_{N_j}=\frac{\pi}{L}N_j(-1)^{2j}
Z_3\left| \begin{array}{ccc}
{\rm g}_1^{(j)}&{\rm g}_2^{(j)}&{\rm g}_3^{(j)}\\
0&0&0
\end{array}\right|\left(-\frac{1}{3}\right).
\label{3.5}
\end{eqnarray}
Taking into account the functional equation (3.4) one gets
\begin{eqnarray}
Z_3\left| \begin{array}{ccc}
{\rm g}_1^{(j)}&{\rm g}_2^{(j)}&{\rm g}_3^{(j)}\\
0&0&0 
\end{array}\right|\left(-\frac{1}{3}\right)
=-\frac{1}{2\pi^3}
Z_3\left| \begin{array}{ccc}
0&0&0 \\
-{\rm g}_1^{(j)}&-{\rm g}_2^{(j)}&-{\rm g}_3^{(j)}
\end{array}\right|\left(\frac{4}{3}\right).
\end{eqnarray}
The Casimir energies (\ref{3.1}) take the form
\begin{eqnarray}
{\cal E}_{N_j}=-\frac{(-1)^{2j}}{2\pi^2L}N_j
Z_3\left| \begin{array}{ccc}
0&0&0 \\
-{\rm g}_1^{(j)}&-{\rm g}_2^{(j)}&-{\rm g}_3^{(j)}
\end{array}\right|\left(\frac{4}{3}\right).
\label{3.7}
\end{eqnarray}
Finally the Casimir energy associated with a multiplet  of fields
characterized by the numbers
$N_0$, $N_{1/2}$, and $N_1$ can be written as follows
\begin{eqnarray}
&&{\cal E}=\sum_j{\cal E}_{N_j}=-\frac{1}{2\pi^2L}
\left[
N_0Z_3\left| \begin{array}{ccc}
0&0&0 \\
-{\rm g}_1^{(0)}&-{\rm g}_2^{(0)}&-{\rm g}_3^{(0)}
\end{array}\right|\left(\frac{4}{3}\right)
\right.\nonumber\\
&&\left.
\quad\mbox{}-
N_{1/2}Z_3\left| \begin{array}{ccc}
0&0&0 \\
-{\rm g}_1^{(1/2)}&-{\rm g}_2^{(1/2)}&-{\rm g}_3^{(1/2)}
\end{array}\right|\left(\frac{4}{3}\right)+
N_{1}Z_3\left| \begin{array}{ccc}
0&0&0 \\
-{\rm g}_1^{(1)}&-{\rm g}_2^{(1)}&-{\rm g}_3^{(1)}
\end{array}\right|\left(\frac{4}{3}\right)
\right].
\label{3.8}
\end{eqnarray} 
Thus, the static Casimir energy for a torus is proportional to $c/L$, 
where $c$ is defined by the features of the GUT under consideration.
Note that the sign of $c$ is {\it a priori\/} unpredictable. 

If for all the fields of the theory we take the same boundary conditions
(periodic or antiperiodic), the $Z_3$'s are all equal, and consequently
the supersymmetry is not broken, and the Casimir energy is zero.
On the other hand, if the different fields satisfy different types of
boundary conditions, supersymmetry is broken and there is a static Casimir
effect.

For the latter situation, consider, as an illustration, the usual case of 
bosons satisfying periodic boundary conditions and fermions satisfying 
antiperiodic boundary conditions on the torus. Then we require only two values.
For the bosons, the Casimir energy is proportional to
\begin{eqnarray}
Z_3\left| \begin{array}{ccc}
0&0&0 \\
0&0&0
\end{array}\right|\left(\frac{4}{3}\right)=16.5323,
\label{3.9}
\end{eqnarray}
which value is given explicitly in Ref.~\cite{zucker}.  For the fermions,
the same reference gives the value
\begin{eqnarray}
Z_3\left| \begin{array}{ccc}
0&0&0 \\
{1\over2}&{1\over2}&{1\over2}
\end{array}\right|\left(\frac{4}{3}\right)=-3.86316.
\label{3.10}
\end{eqnarray}
So each term in Eq.~(\ref{3.8}) contributes a negative energy.  
Thus the net Casimir energy is attractive,
\begin{eqnarray}
{\cal E}=-{c\over L}, \quad c=0.837537(N_0+N_1)+0.195710N_{1/2}
=1.033247N_{1/2},
\label{realcas}
\end{eqnarray}
since the number of fermions must be equal to the number of 
bosons.\footnote{Any case with periodic boundary conditions in some 
directions and antiperiodic ones in others may be given in terms of the
values given in Eqs.~(\protect\ref{3.9}), 
(\protect\ref{3.10}) and the additional values
$$
Z_3\left|\protect\begin{array}{ccc}
0&0&0\\
{1\over2}&0&0\protect\end{array}\right|\left(\frac{4}{3}\right)=0.689223,\quad
Z_3\left|\protect\begin{array}{ccc}
0&0&0\\
{1\over2}&{1\over2}&0\protect\end{array}\right|
\left(\frac{4}{3}\right)=-2.156887.
$$}

We should make the following general remarks concerning the physical
interpretation
of the  calculation sketched here. Imposition of periodic boundary
conditions at the boundaries of the field volume is a basic physical ingredient
 in expressions such as Eqs.~(\ref{3.1}), (\ref{3.2}) for the Casimir energy. 
 It is analogous to the
imposition of perfect conducting boundary conditions, or more generally,
electromagnetic boundary conditions, at the walls, when considering
ordinary electrodynamics, for example within a spherical volume.
The physical outcome of a calculation of this
kind is the residual energy  remaining when the influence of the
local stresses is separated off.
(Presumably, such stresses are absorbed in a kind of renormalization
of physical parameters.) The field theoretical calculation
is able to cope only with the cutoff independent part of the physical stress;
the local cutoff dependent parts of the stress are 
automatically lost in the zeta-function regularization
process. This is an important point
whenever the result of the field theoretical calculation is to be compared
with experiments.

As a typical example of this sort, we may mention the calculation of the
Casimir energy of a dilute dielectric  ball. One may adopt a field
theoretical viewpoint (cf., for instance, \cite{brevik99a}), from which the
Casimir energy is calculated as a cutoff independent, positive, expression.
More detailed considerations, using quantum mechanical perturbation theory
\cite{barton99} (cf.~also \cite{bordag99}), or quantum statistical mechanics
\cite{hoye99} show however how this expression  is to be supplemented with
attractive cutoff dependent parts. Such terms are presumably not observable.
As for the cutoff independent term,
agreement between the methods is found, so the situation is in this respect
satisfactory.

\section{Dynamical properties}

Now let us turn to a simplified discussion of the dynamical Casimir effect.
We here take into account that we have a
dynamical radius $a(t)L$, $a(t)$ being a dimensionless scale factor.
 Then, the total effective action is given as
\begin{eqnarray}
\Gamma=W-L\int d \eta \,a(\eta)\,{\cal E},
\label{4.1}
\end{eqnarray}
where $W$ is given by Eq.~(\ref{2.4}) and ${\cal E}=-c/(aL)$ as displayed
in Eq.~(\ref{3.8}). Because the action is dimensionless, the
length $L$ disappears from the calculation, and we have
\begin{eqnarray}
\Gamma=\int d \eta \left[ 2b' \sigma
\sigma''''-2(b+b')(\sigma''+\sigma^{\prime2})^2+ c \right].
\label{4.2}
\end{eqnarray}
This is a typical effective action to describe a quantum FRW Universe.

In order to consider dynamical properties we add to the above effective action
a phenomenological term, which has the form of a kinetic energy.
We  associate a mass $m$ with the scale factor $a$, and take the
corresponding kinetic
energy to be given by $\frac{1}{2}m\dot{a}^2$.  Our essential idea is that the
geometrical configuration of the space is treated classically and that the GUT
field is a quantum object which induces the Casimir effect.  One might in
principle introduce other expressions for the kinetic energy, but this
expression is clearly 
the simplest choice that one can make. The Newtonian form is
moreover in correspondence with our use of the adiabatic approximation,
meaning that $|\dot{a}(t)| \ll 1$; cf. also the analogous argument in
\cite{nagatani99} in connection with the (1+1) dimensional  case.\footnote{We
consider other possibilities for the kinetic energy term in the Appendix.}
Introducing the physical time $t$
via $d \eta /dt=1/a$, we now write $\Gamma$ as
\begin{eqnarray}
\Gamma &=& \int dt\,\left[
\frac{1}{2}m\dot{a}^2+2b' \ln a\,(\stackrel{....}{a}
a^2+3\stackrel{...}{a}\dot{a}a
       + \ddot{a}^2a+\ddot{a}\dot{a}^2)\right.
\nonumber\\
&&\left.\mbox{}-2(b+b')\frac{(\dot{a}^2+a\ddot{a})^2}{a}
+\frac{c}{a}
\right],
\label{4.3}
\end{eqnarray}
where $\dot{a}=da/dt$.

\mbox{}From the variational equation $\delta \Gamma/\delta a =0$ we obtain, 
after some algebra,\footnote{This and subsequent equations are dimensionally
consistent if we restore dimensions: 
$$[a]=[t]=[m^{-1}]=\mbox{Length}.
$$  }
\begin{eqnarray}
&&m\ddot{a}
-2\,b'\,\left(
\ddot{a}^2+2\frac{d^2}{dt^2}(a\ddot{a} )
\right)
\nonumber\\
&&\mbox{}-2\,(b+b')\,\left(
2a\stackrel{....}{a}+4\dot{a}\stackrel{...}{a}+3\ddot{a}^2
-12\frac{\dot{a}^2\ddot{a}}{a}+3\frac{\dot{a}^4}{a^2}
\right)
+\frac{c}{a^2}=0.
\label{4.4}
\end{eqnarray}
It is remarkable that the the logarithm is absent in Eq.~(\ref{4.4}); 
there seems to be no reason {\it a priori} why this should be so.

We limit ourselves to the ${\cal N}=4$ $SU(N)$ super YM theory for which, as
mentioned, $(b+b')=0$. Then Eq. (4.4) simplifies to
\begin{eqnarray}
m\ddot{a}+2b(2\stackrel{....}{a}a+4\stackrel{...}{a}\dot{a}+3\ddot{a}^2)+
\frac{c}{a^2}=0.
\label{4.5}
\end{eqnarray}
Both the terms involving $b$ and $c$ are dynamical, quantum mechanical, 
effects, which in
dimensional terms are proportional to $\hbar$. However, we will
see that it is sensible (if $c\ne0$) to regard the $b$ term as a
 small correction to the Casimir-determined geometry. We denote the
$b=0$ solution by $a_0(t)$; it
satisfies the equation
\begin{eqnarray}
m\ddot{a}_0 +\frac{c}{a_0^2}=0,
\label{4.6}
\end{eqnarray}
implying
\begin{eqnarray}
\frac{1}{2}m\dot{a}_0^2=\frac{c}{a_0}+\mbox{const}.
\label{4.7}
\end{eqnarray}

\subsection{Attractive Casimir energy, $c>0$}
If the Casimir energy is attractive, as actually realized in our 
illustrative calculation given in Sec.~3, see Eq.~(\ref{realcas}),
 we will assume, 
as boundary conditions, that $a_0(t \rightarrow \infty)=\infty$,
$\dot{a}_0(t\rightarrow \infty)=0$. Then, the constant in Eq.~(\ref{4.7}) becomes
equal to zero, and we get the Casimir solution
\begin{eqnarray}
a_0(t)=A t^{\frac{2}{3}},~~~~{\rm with}~~~~ A=\left( \frac{9c}{2m}
\right)^{\frac{1}{3}}.
\label{4.8}
\end{eqnarray}
It is worth noticing here that the proportionality of $a_0(t)$ to $t^{2/3}$ is
precisely the behaviour shown by the scale factor in the Einstein-de Sitter
universe. This may be surprising at first sight, but does not seem to be so
unreasonable after all, since the Einstein-de Sitter universe is flat, thus
in correspondence with our neglect of Riemannian curvature terms in the
formalism above.

Now we turn to the solution $a(t)$, taking into account the $b$
correction. We
shall limit ourselves to giving a perturbative solution, implying an 
expansion of $a(t)$ around $a_0(t)$ assuming $b$ to be small:
\begin{eqnarray}
a(t)=a_0(t)+b a_1(t).
\label{4.9}
\end{eqnarray}
We consider only times for which the correction term is small:
\begin{eqnarray}
ba_1/a_0 \ll 1.
\label{4.10}
\end{eqnarray}
Thus, we may expand the Casimir term in Eq.~(\ref{4.5}) as  $c/a^2=
(c/a_0^2)(1-2ba_1/a_0)$. A first order expansion of  the other terms in
Eq.~(\ref{4.5}) 
then yields, when we take into account the Casimir solution (\ref{4.8}), the
inhomogeneous equation
\begin{eqnarray}
\ddot{a}_1-\frac{4}{9t^2}a_1=\frac{8A^2}{9m} t^{-\frac{8}{3}}.
\label{4.11}
\end{eqnarray}
The homogeneous version of Eq.~(\ref{4.11}) 
has solutions of the form $t^{\alpha}$,
with $\alpha =4/3$ and $\alpha = -1/3$.  We write the independent solutions as
\begin{eqnarray}
f(t)=t^{\frac{4}{3}}, \quad g(t)=t^{-\frac{1}{3}}.
\label{4.12}
\end{eqnarray}
The Wronskian $\Delta$ between $f$ and $g$ is simple;
$\Delta=f\dot{g}-g\dot{f}=-5/3$. Writing for brevity the right hand side of
Eq.~(\ref{4.11}) as $r$ we then get, as 
the solution of the inhomogeneous equation,
\begin{eqnarray}
a_1(t)&=&f(t) \left( C_1-\frac{1}{\Delta}\int rg \,
dt\right)+g(t)\left(C_2+\frac{1}{\Delta}\int rf\, dt \right)\nonumber\\
&=&t^{\frac{4}{3}}\left(C_1-\frac{4A^2}{15m}t^{-2}
\right)+t^{-\frac{1}{3}}\left(
C_2+\frac{8A^2}{5m}
t^{-\frac{1}{3}}\right)\nonumber\\
&=&C_1t^{4/3}+C_2t^{-1/3}+{4\over3}{A^2\over m}t^{-2/3},
\label{4.13}
\end{eqnarray}
with $C_1$ and $C_2$ being constants.
  As for the values of these
constants, we
have first to observe our restriction (4.10), which implies that
\begin{eqnarray}
\frac{b}{A}\left(
C_1t^{\frac{2}{3}}+C_2t^{-1}+\frac{4A^2}{3m}t^{-\frac{4}{3}}\right) \ll 1.
\label{4.14}
\end{eqnarray}
If we require the perturbative approximation to be valid for large times,
we must have $C_1=0$. If we also set $C_2=0$, our perturbative 
solution becomes
\begin{eqnarray}
a(t)=At^{\frac{2}{3}}+\frac{4A^2}{3m}bt^{-\frac{2}{3}},
\label{4.15}
\end{eqnarray}
which is only valid for large enough $t$, i.e., for
\begin{eqnarray}
\frac{bA}{m}t^{-\frac{4}{3}} \ll 1.
\label{4.16}
\end{eqnarray}
The static Casimir force is
\begin{eqnarray}
F_{\rm Cas}=-\frac{\partial}{\partial a}\left(-\frac{c}{a}\right)=-\frac{c}{a^2},
\label{4.17}
\end{eqnarray}
whereas the dynamical force is
\begin{eqnarray}
F_{\rm dyn}=m\ddot{a}.
\label{4.18}
\end{eqnarray}
Substituting Eq.~(\ref{4.15}) into Eqs.~(\ref{4.17}) and (\ref{4.18}), and
observing the relation between $c$ and $A$ in Eq.~(\ref{4.8}), we get
\begin{eqnarray}
F_{\rm dyn}=F_{\rm Cas}\left(1-\frac{4bA}{m}t^{-4/3}\right),
\end{eqnarray}
which shows that the dynamical force is the Casimir force modified by
a small dynamical correction when the perturbative approximation is valid.

Before we turn to a numerical solution of Eq.~(\ref{4.5}), we discuss the
repulsive case.

\subsection{Repulsive Casimir energy, $c<0$}

Now let us consider the case when $c<0$, a repulsive Casimir energy. 
In this case we must take
$\dot{a}_0|_{t \to \infty}\ne 0$. Let us write the $b=0$ equation (\ref{4.7}) 
in the form
\begin{equation}
\label{5.1}
\dot{a}_0=\pm\sqrt{\frac{c_1}{a_0}+c_2},
\end{equation}
\begin{equation}
c_1=\frac{2c}{m},
\qquad
c_2=v_{\infty}{}^2,
\qquad
v_{\infty}=\dot{a}_0|_{t \to \infty}.
\end{equation}
\mbox{}From (\ref{5.1}) we obtain
\begin{equation}
\label{5.2}
{1\over c_2}\left[a_0\sqrt{\frac{c_1}{a_0}+c_2}-\frac{c_1}{\sqrt{c_2}}
\ln\left( 2\sqrt{c_2 a_0}+2\sqrt{c_2a_0+c_1} \right)\right] =\pm t + c_3,
\end{equation}
where $c_3$ is a further integration constant.
We see from Eq.~(\ref{5.1}) that
\begin{equation}
\label{5.3}
a_0\ge \frac{-2c}{mv_{\infty}^2}.
\end{equation}
For long times, the solution behaves as
\begin{eqnarray}
a_0(t)\sim\sqrt{c_2}t,\quad t\gg1.
\label{larget}
\end{eqnarray}
For short times, suppose $a_0$ approaches the minimum value (\ref{5.3}); then
\begin{eqnarray}
c_3=-{c_1\over c_2^{3/2}}\ln2\sqrt{-c_1},
\end{eqnarray}
and
\begin{eqnarray}
a_0(t)\sim-{c_1\over c_2}-{c_2^2\over4c_1}t^2,\quad t\ll1.
\label{smallt}
\end{eqnarray}
\section{Numerical solution and discussion}

\subsection{$c>0$}
Let us consider numerical solutions of dynamical equations (\ref{4.5}) 
for $SU(2)$  super Yang-Mills theory, for the attractive case.
We suppose that the initial behavior of $a(t)$ is given by the perturbative
form Eq.~(\ref{4.15}). We may always set $m=1$ since that amounts to using
dimensionless variables for $a$ and $t$. 
Let us take as an illustration
\begin{eqnarray}
N=2,
\qquad
c=1\quad (A=1.65096),
\qquad
t_0=0.5,
\end{eqnarray}
For later times we integrate the exact equations numerically, starting with the
initial conditions at $t_0$:
\begin{eqnarray}
a_0=1.06744,
\quad\dot{a}_0=1.35019,
\quad\ddot{a}_0=-0.802706,
\quad \overdots{a}_0=1.81582.
\end{eqnarray}
For those conditions we have a numerical solution for $a(t)$ as shown in
Fig.~\ref{fig1}.  For comparison we also show in 
the figure the unperturbed solution (\ref{4.8}) due 
to the static Casimir force.  
It will be noticed that for large $t$ there are significant deviations
from the unperturbed solution, which must be due to $C_1\ne0$ in the
perturbative solution (\ref{4.13}).  In fact, for the entire range of
$t=0.1$--80, the exact solution shown in Fig.~\ref{fig1} is roughly
reproduced by Eq.~(\ref{4.13}) with $C_1=C_2=-5$.
For the Casimir force we have the behavior as shown in Fig.~\ref{fig2}.
The exact solution has oscillations, but overall is 
close to the unperturbed solution.
Not surprisingly, the Casimir energy dominates the force.
Note also that for another choice of initial conditions one
will find somewhat different behaviour.  The essential
property of the approximation
under discussion is that there are always dynamical oscillations around the
static Casimir force.

\subsection{$c<0$}

Next we consider numerical solutions of dynamical equations (\ref{4.5})
for $SU(2)$  super Yang-Mills theory when $c<0$.
We suppose that the initial behavior of $a(t)$ is given by
form Eq.~(\ref{5.2}). Let us take the illustrative values
\begin{eqnarray}
N=2,
\quad
c_1=-1,
\quad
c_2=1,
\quad
a_0|_{t=0}=1,
\quad
t_0=0.1,
\end{eqnarray}
For later times we integrate the exact equations numerically, starting
from the initial conditions at $t_0$:
\begin{eqnarray}
a_0=1.00241,
\quad
\dot{a}_0=0.0490327,
\quad
\ddot{a}_0=0.497511,
\quad
\stackrel{...}{a}_0\,=-0.049793.
\end{eqnarray}
Note that the perturbative values of these parameters, given from
Eq.~(\ref{smallt}) are close to these:
\begin{eqnarray}
a_0=1.0025,
\quad
\dot{a}_0=0.05,
\quad
\ddot{a}_0=0.5,
\quad
\stackrel{...}{a}_0\,=0.
\end{eqnarray}
For those conditions we have a numerical solution for $a(t)$ as shown in
Fig.~\ref{fig3}.
 For the Casimir force we have the behavior as shown in
Fig.~\ref{fig4}.  For both cases the exact solution has
very small oscillations, but overall is close to the Casimir solution,
and is accurately described by the limits of that solution, Eqs.~(\ref{smallt})
and (\ref{larget}).

Thus, we have presented a formalism to describe the dynamical Casimir effect in the
adiabatic approximation. It may be applied to an arbitrary GUT. Without any
technical problems one can generalize the present consideration to any
specific 
four-dimensional background (we limited ourselves to a discussion of a toroidal
FRW universe as providing the moving boundary conditions). But the limitations
of our approach must be stressed: It would be extremely interesting to
suggest new formulations of the dynamical Casimir effect beyond the adiabatic
approximation.


\section*{Acknowledgments}
 We thank A. Bytsenko for helpful discussions. 
The work by SDO and KEO has been supported in part by RFBR,
that of SDO also by CONACyT, and that of KAM by the US Department of Energy.

\appendix
\section{Alternative Kinetic Energy Terms}

In the text, we introduced an {\it ad hoc\/} kinetic energy term into
the action, referring to the change of scale with physical time,
\begin{equation}
\int dt\,{1\over2}m\dot a^2.
\label{simpleke}
\end{equation}
This is rather natural in the adiabatic context, where $|\dot a|\ll1$,
for then simple scaling properties obtain, as evidenced by the dimensional
consistency of the resulting equations of motion when $m$ has dimensions
of mass.  But we can not offer very strong arguments in its favor, in the
absence of dynamical information.  So, in this appendix we consider two
alternatives, which provide somewhat different models for the dynamical
evolution of the world.

In the first, we suppose that the same kinetic energy should be
integrated over conformal time,\footnote{In
this case the parameter $m$ is dimensionless.}
\begin{equation}
\int d\eta\, {1\over2}m\dot a^2=m\int dt\,{1\over2}{\dot a^2\over a},
\label{mt1}
\end{equation}
so that the Casimir evolution equation is, in place of Eq.~(4.7),
\begin{equation}
{1\over2}{m \dot a^2\over a}={c\over a}+k,
\end{equation}
where we have dropped the subscript 0 for simplicity, and written the constant
of integration as $k$.  The solution of this equation is very simple,
\begin{equation}
a=a_0+t\sqrt{2\over m}\sqrt{c+ka_0}+{k\over2m}t^2,\quad a_0=a(0).
\label{A4}
\end{equation}
If $c>0$, we can set $k=0$ and obtain instead of the behavior
exhibited in Eq.~(4.8), a linear growth of the scale,
\begin{equation}
a=a_0+\sqrt{2c\over m} t.
\end{equation}
If $c<0$, as before we cannot set the integration constant equal to zero;
if we again choose the initial velocity to be zero, or $a_0=-c/k$,
we get a result very like Eq.~(4.26):
\begin{equation}
a=-{c\over k}+{k\over2m}t^2,
\end{equation}
but now valid for all times.

Perhaps a more natural possibility is to use the conformal time everywhere
in the kinetic energy term,\footnote{In this case the parameter $m$ has
dimension 1/Length$^2$.}
\begin{equation}
\int d\eta \,{m\over 2}\left(da\over d\eta\right)^2={m\over2}\int dt 
\,a\,\dot a^2.
\label{mt2}
\end{equation}
The solution to the purely Casimir dynamical equation is
\begin{equation}
{m\over2}a\dot a^2={c\over a}+k,
\end{equation}
which is integrated to
\begin{equation}
t={1\over k^2}\sqrt{m\over2}\left\{{2\over3}\left[(ka+c)^{3/2}-(ka_0+c)^{3/2}
\right]-2c\left[(ka+c)^{1/2}-(ka_0+c)^{1/2}\right]\right\}.
\label{A9}
\end{equation}
When $c>0$ again we can take $k\to0$, which leads to the $k=0$ result
\begin{equation}
a^2=a_0^2+2\sqrt{2c\over m}t.
\end{equation}
If $c<0$ and we choose again $ka_0+c=0$, we obtain for short times
\begin{equation}
a=-{c\over k}+{k^3\over 2mc^2}t^2, \quad t\ll 1,
\end{equation}
again very similar to Eq.~(4.26).

In Figs.~\ref{figapp} and 
\ref{figapp2} we show the effect of the inclusion of the dynamical $b$
term for these kinetic energy structures. Qualitatively, the results
do not depend much on whether the Casimir term is positive or negative.
The example given for the first alternative 
kinetic energy is similar to the simple model result 
for $c<0$ shown in Fig.~\ref{fig3} except that the growth in $t$
is quadratic rather than linear.
The evolution for the 
second form of kinetic energy resembles the simple model result
for $c>0$ shown in Fig.~\ref{fig1}.

\newpage

\begin{figure}
\centerline{\psfig{figure=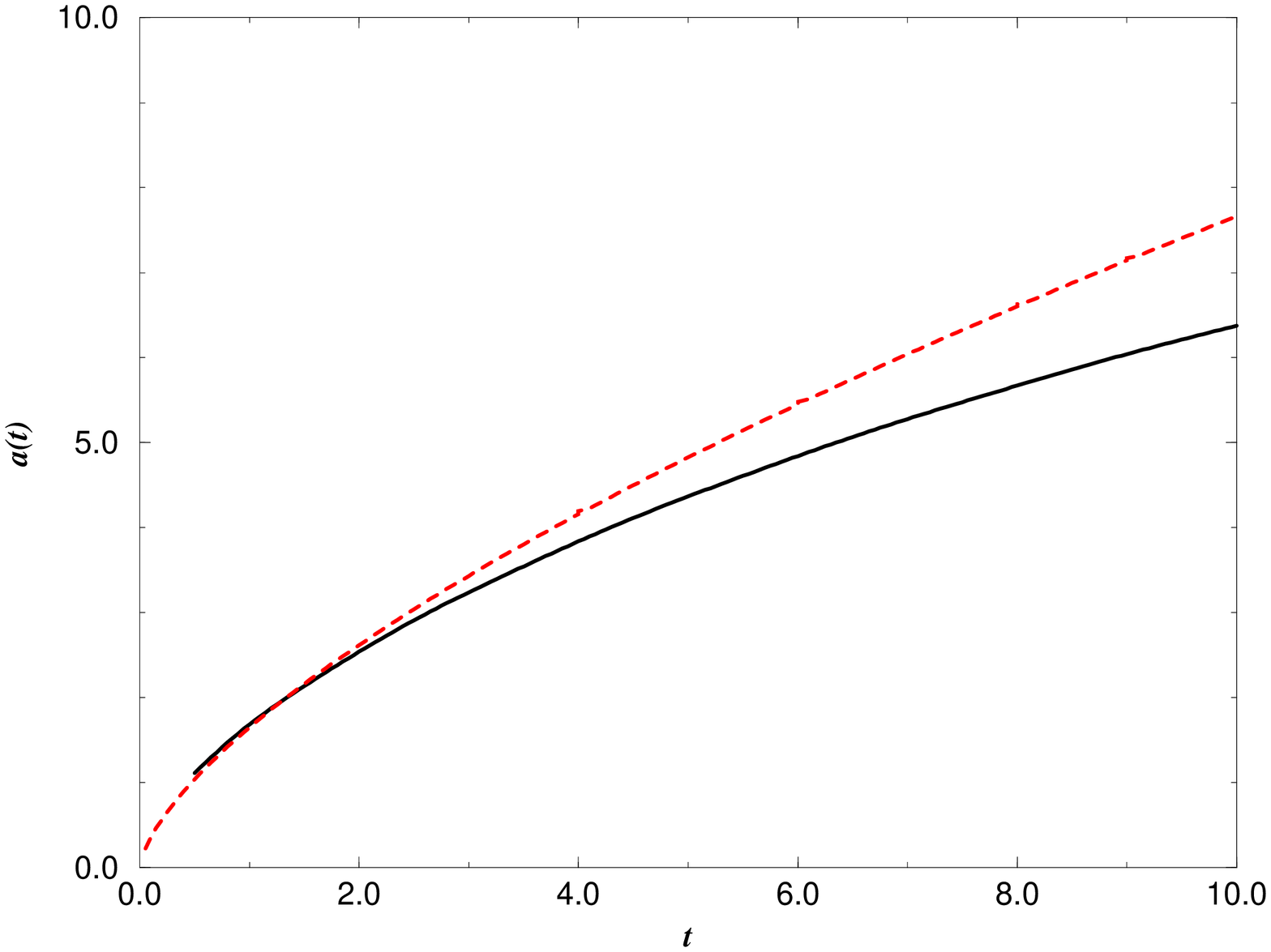,height=45ex,angle=270}}
\caption{Casimir (dashed line) and
dynamical behavior for $a(t)$ for $c>0$.}
\label{fig1}
\end{figure}
\begin{figure}
\centerline{\psfig{figure=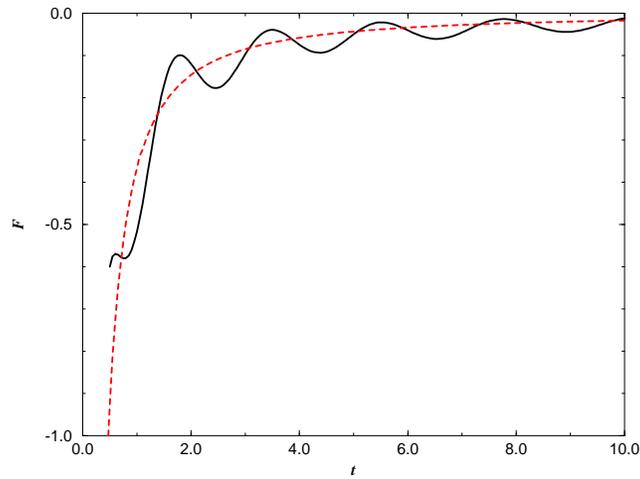,height=45ex,angle=270}}
\caption{Casimir (dashed
line) and dynamical behavior of $F$ for $c>0$.}
\label{fig2}
\end{figure}
\begin{figure}
\centerline{\psfig{figure=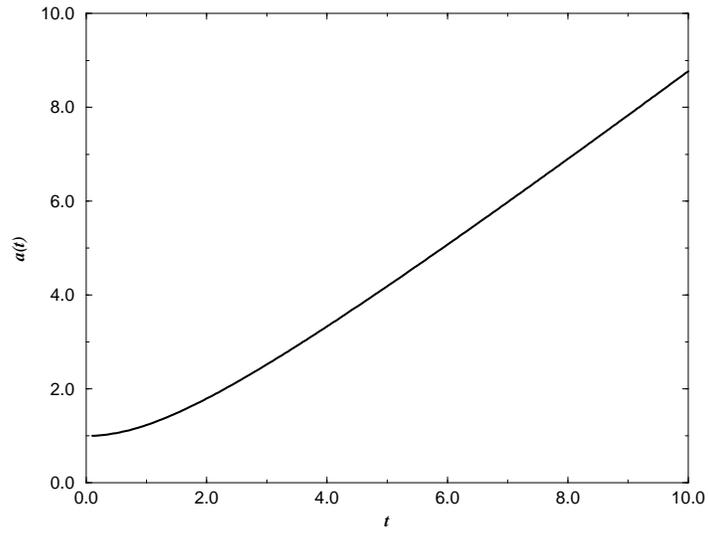,height=50ex,angle=270}}
\caption{Dynamical behavior of $a(t)$ for $c<0$.}
\label{fig3}
\end{figure}
\begin{figure}
\centerline{\psfig{figure=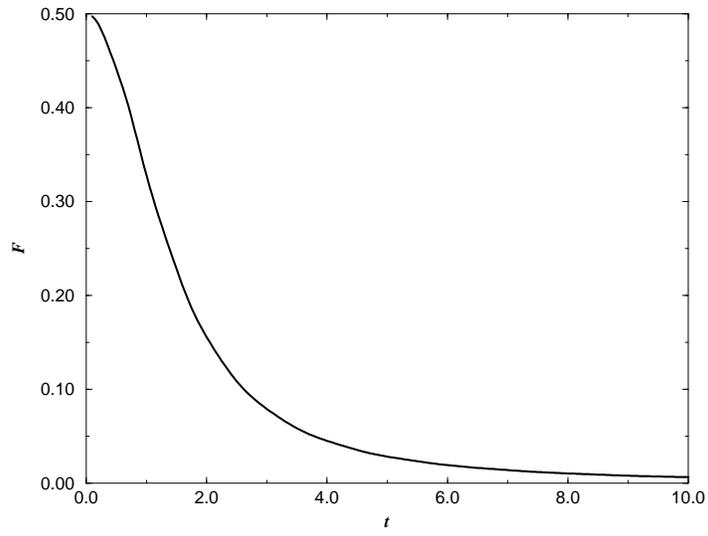,height=50ex,angle=270}}
\caption{Dynamical behavior of $F$ for
$c<0$.}
\label{fig4}
\end{figure}

\begin{figure}
\centerline{\psfig{figure=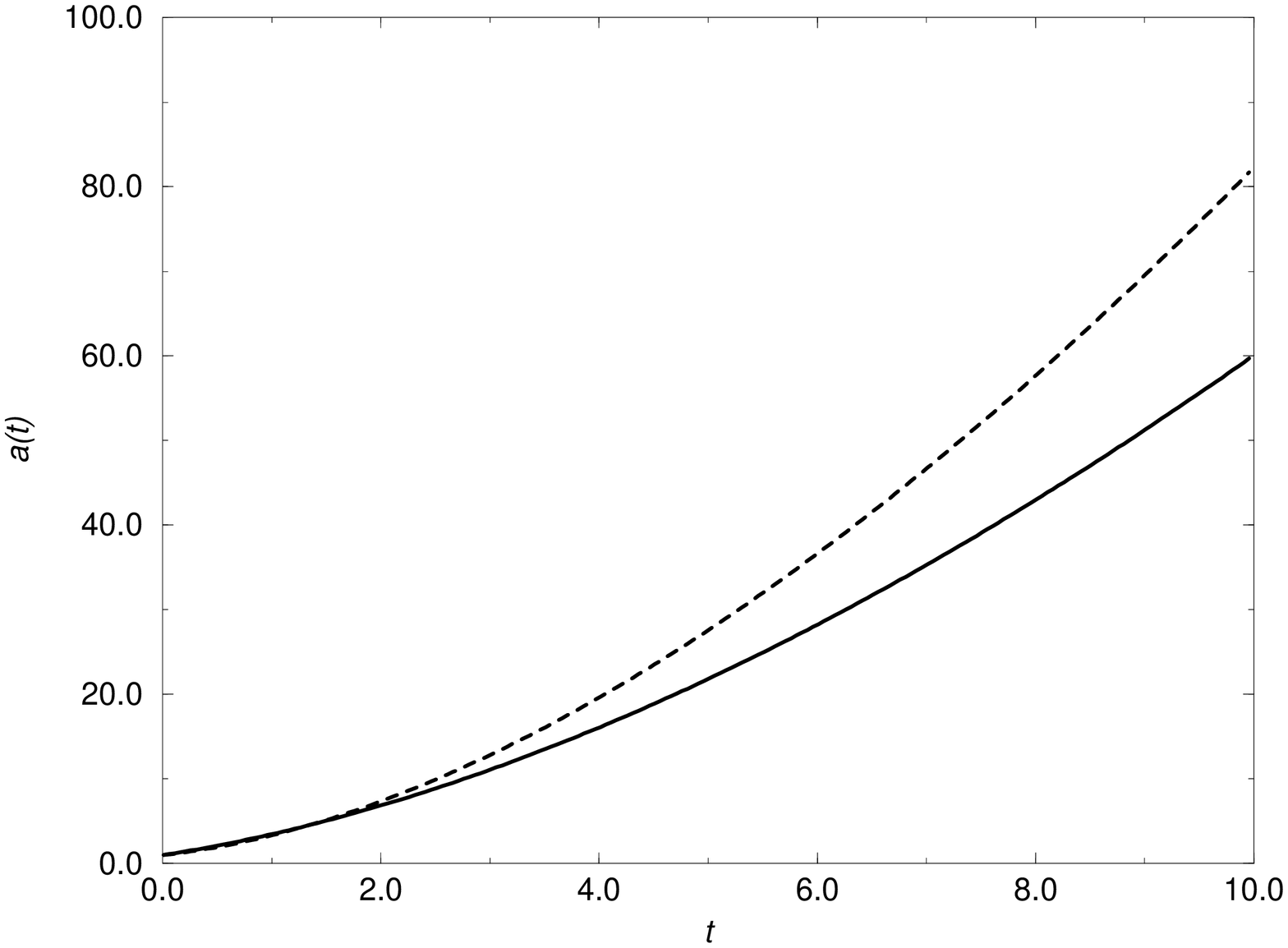,height=45ex,angle=270}}
\caption{Dynamical behavior for $a(t)$ for the first alternative
kinetic energy term, Eq.~(\protect\ref{mt1}).  Shown are the behaviors
with $m=1$, and initial condition $a(0)=1$, evolving initially 
until $t=0.01$ according
to Eq.~(\protect\ref{A4}), with $c=1$, $k=1$ (solid line); and
with $c=-1$, $k=2$ (dashed line).}
\label{figapp}
\end{figure}

\begin{figure}
\centerline{\psfig{figure=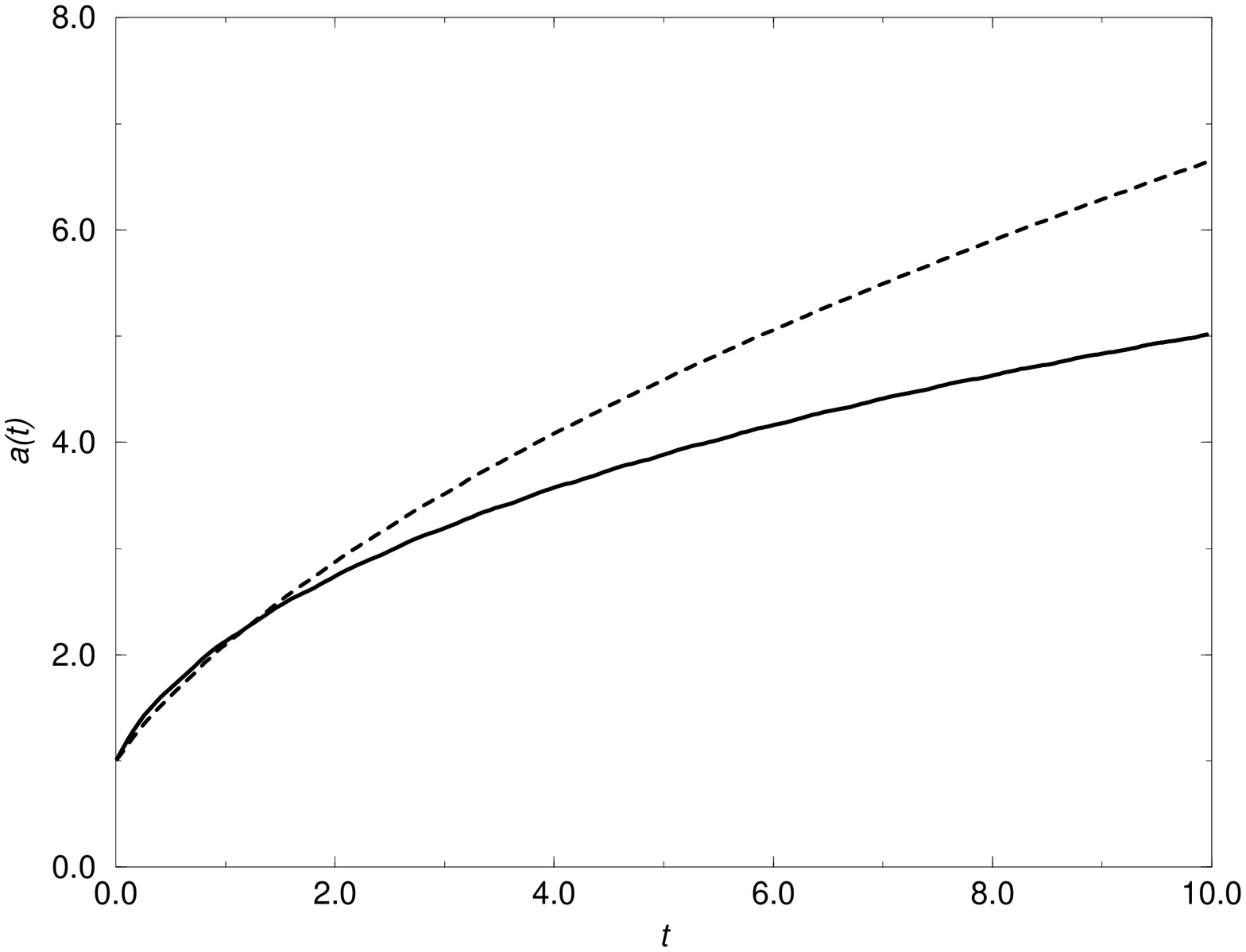,height=45ex,angle=270}}
\caption{Dynamical behavior for $a(t)$ for the second alternative
kinetic energy term, Eq.~(\protect\ref{mt2}).  Shown are the behaviors
with $m=1$, and initial condition $a(0)=1$, evolving initially
until $t=0.01$ according
to Eq.~(\protect\ref{A9}), with $c=1$, $k=1$ (solid line); and
with $c=-1$, $k=2$ (dashed line).}
\label{figapp2}
\end{figure}

\end{document}